\documentclass[twocolumn,10pt,a4paper,float,
amsmath, amssymb, aps,
prl,
]{revtex4-2}

\usepackage[utf8]{inputenc}

\usepackage{float}
\usepackage{graphicx}
\usepackage{dcolumn}
\usepackage{bm}
\usepackage{mathrsfs}

\newcommand{\la}{\langle}
\newcommand{\ra}{\rangle}
\newcommand{\nex}{n_{\text{ex}}}

\usepackage[dvipsnames]{xcolor}
\usepackage{hyperref}
\hypersetup{
    colorlinks=true,
    citecolor=BrickRed,
    linkcolor=RoyalBlue
}

\begin{document}

\title{\large Quantum supremacy of the many-body fluctuations \\in the occupations of the excited particle states in a Bose-Einstein-condensed gas}

\author{V.V. Kocharovsky$^1$, Vl.V. Kocharovsky$^2$, and S.V. Tarasov$^2$\\
\textit{$^{1}$Department of Physics and Astronomy, Texas A\&M University, College Station, TX 77843, USA}\\
\textit{$^{2}$Institute of Applied Physics, Russian Academy of Sciences, Nizhny Novgorod 603950, Russia}
}
\date{\today}

\begin{abstract}
We find a universal analytic formula for a characteristic function (Fourier transform) of a joint probability distribution for the particle occupation numbers in a BEC gas and the Hafnian Master Theorem generalizing the famous Permanent Master Theorem of MacMahon.
We suggest an appealing model, a multi-qubit BEC trap formed by a set of qubit potential wells, and discuss specifics of such an atomic boson-sampling system vs a photonic one. 
Finally, the process of many-body fluctuations in a BEC trap is $\sharp$P-hard for computing. 
It could serve as a basis for demonstrating quantum advantage of the many-body interacting systems over classical simulators.
\end{abstract}

\maketitle
	
{\it Many-body BEC fluctuations as a quantum simulator for boson sampling.} -- A concept of quantum advantage of many-body quantum simulators over classical computers is in the spotlight of modern quantum physics \cite{Harrow2017,Boixo2018,Arute2019,Zhong2020,Dalzell2020}. 
For its testing, a boson sampling of the single-photon Fock states in a linear interferometer had been suggested in \cite{Aaronson2011,Aaronson2013}. 
Yet, an absence of suitable on-demand sources of single photons put forward the boson sampling of Gaussian, squeezed states of photons as the most plausible platform \cite{Aaronson2013,Scheel2004,LundPRL2014,Bentivegna2015,Shi2021,Kalai2016,Wu2016,ShchesnovichPRL2016,Shchesnovich2019,Wang2017,He2017,Loredo2017,Hamilton2017,Hamilton2019,Chin2018,Quesada2018,Zhong2019,Paesani2019,Brod2019,Huh2019,Huh2020,DrummondReid2019,Drummond2021,Wang2019,PanPRL2021,Villalonga2021}. 
We consider an alternative platform based on the Bose-Einstein condensate (BEC) of trapped atoms.
The starting point of our analysis is a fact of two-mode squeezing of particle excitations in a BEC trap established in \cite{PRA2000} and strongly pronounced in the fluctuations of a total BEC occupation calculated in \cite{PRA2020}. 

Physics of $N$ atoms in a BEC trap looks substantially different from the physics of massless photons in the interaction-free, nonequilibrium (nonthermal), linear interferometer due to the presence of the condensate, thermal equilibrium, particle mass and interaction as well as the absence of external sources of particles.
Still, we show that these peculiarities turn the BEC trap into a platform for observing the boson-sampling quantum advantage. 

Consider joint fluctuations in the occupations of the excited particle states.
We find a truly simple, universal formula for their characteristic function (Fourier transform of their joint probability distribution) in terms of a normally-ordered correlation function $G$ of trapped particles. 
By the MacMahon Master Theorem \cite{MacMahon1916,Percus1971} and the Hafnian Master Theorem (\ref{HMT}), it yields the cumulants (hence, moments) and probabilities of the joint distribution via a matrix permanent and a hafnian (a certain extension of the permanent \cite{Barvinok2016}) which are $\sharp$P-hard to compute \cite{Valiant1979,Jerrum2004} and viewed as a universal tool for analyzing the $\sharp$P-hard problems \cite{Entropy2020,Harrow2017}. 
This fact justifies a quantum advantage of many-body equilibrium fluctuations in the occupations of excited particle states in a BEC trap and opens a path for the exploration of an entire spectrum of the theoretical/experimental BEC problems inspired by boson sampling in an interferometer. 

For simplicity's sake, we consider an equilibrium BEC at temperatures well below the critical region, within the Bogoliubov-Popov approximation \cite{Popov1965,Shi1998}. 
We show that computing particle excitation fluctuations is still a $\sharp$P-hard problem (even within the grand canonical ensemble \cite{footnote1,JStatPhys2015,PLA2015,PhysScr2015,arXivIsing2016}). 
This is true if there are (a) interparticle interactions and (b) nonuniformity of the condensate leading to Bogoliubov coupling between a sufficiently large number of the excited particle states \cite{footnote2x2}. 

Experimental studies of BEC in dilute gases \cite{Kristensen2019,MehboudiPRL2019,Hadzibabic2017QD,ChangPRL2016,Dalibard2015,Perrin2012,toroidBEC-PRL2011,Hadzibabic2010,Armijo2010,Jacqmin2010,Campbell,Hung2011,Cornell2010,Dalibard2008,Esslinger2007} had allowed one to directly measure fluctuations in a total BEC occupation. 
Measuring occupations of individual excited states will come soon. 
Their understanding means reaching a much deeper level of quantum statistical than a level of the mean condensate, quasiparticle characteristics and condensate fluctuations studied previously \cite{PitString2016,Steinhauer2002,Kuklov2006,Makotyn2014,ChangPRL2016,Hadzibabic2017QD,Hadzibabic2019excitations,Pieczarka2020}. 
Particle-number fluctuations are important for matter-wave interferometers \cite{Shin2004} (like Ramsey \cite{DrummondPRA2019,Drummond2011} or Mach-Zehnder \cite{Chip1000atoms} on-chip ones) and were studied for squeezed states \cite{DrummondPRA2019}, trap cells \cite{Castin,Pit2011}.

{\it A potential trap design featuring quantum advantage: The BEC trap made up of the qubit potential wells.} -- As is shown below, general-case BEC traps have quantum advantage over classical simulators. 
To demonstrate this advantage in a controllable and clear way one could use specially designed traps. The challenge is twofold. 
First, a trap with a finite number $M$ of relatively well populated excited stated, coupled to each other via Bogoliubov coupling, is desirable. 
If all higher excited states are separated from such a lower miniband by an energy gap wider than the temperature $T$, they would have exponentially small occupations and can be skipped or accounted for as a kind of perturbation. 
Second, there should be a way to sample, simultaneously measure the occupations of the lower miniband states, say, via a multi-detector imaging. 

Consider a multi-qubit design of the BEC trap with a split-off lower energy miniband: Separate several, $Q$, tight qubit cells, each with two close lower energy levels, \cite{footnoteTrapSize} by relatively narrow, not very high potential barriers and arrange them in a two- or three-dimensional lattice (see figure in \cite{SM}). Place the lattice on top of a slightly varying in space background potential with high walls at the trap borders. Quantum tunneling of atoms under the inter-cell barriers should be significant to ensure a reasonable interaction between atoms from different individual qubit wells needed for a formation of a common nonuniform condensate and significant Bogoliubov couplings within a large subset of excited states (for another design, see \cite{footnote1D}). Otherwise, the boson sampling would simplify and lose its quantum advantage. Adjust parameters to form a lower miniband of $M+1=2^Q$ levels separated from all higher levels by an energy gap $\Delta E > T$.  

An individual qubit well has a twofold-degenerate ground level split by a certain perturbation. 
In particular, a double-well trap becomes the qubit well if its parameters are adjusted appropriately. 
BEC in the double-well traps and optical lattices, their Bogoliubov excitations are well studied \cite{PitString2016,Salasnich1999,Ho2000,Shin2004,Gati2006,Baym2006,DrummondPRA2019,Masiello2019,Pokrovsky2020,Okeke2021}.

Lowering the temperature below the critical value $T_c$ and controlling the inhomogeneous background potential and barriers separating qubit wells allow one to create an entire hierarchy of BEC regimes \cite{Baym2006}: 
From the regime of anomalously large critical fluctuations in the critical region (neat $T_c$) or strongly correlated regime to the regime of a quasi-condensate or fragmented condensates of the individual qubit wells to the regime of a well established, macroscopically occupied common condensate inhomogeneously spread over the entire trap at $T \ll T_c$. 
We consider the latter case assuming $N \gg Q$ \cite{footnoteDecoupledMiniband}. 

{\it Joint probability distribution of the excited particle occupations via the characteristic function, the Hafnian Master Theorem.} -- Quantum transitions of particles between excited states are described by the operators $\hat{a}_k^{\dagger}$ and $\hat{a}_k$ which create and annihilate, respectively, a particle in a state with a wavefunction $\psi_k(\bf{r})$ in a mesoscopic trap of a finite volume $V$ confining a dilute interacting gas of $N$ particles in total by means of some external potential $U(\bf{r})$. 
Let us consider an equilibrium state (described by a density matrix $\hat{\rho}$) of such a Bose-Enstein-condensed gas with a well-formed macroscopic wave function $\psi_0(\bf{r})$ of the condensate at a temperature $T$ below a critical region.
This $N$-body system can be accurately described by means of the Bogoliubov-Popov approximation \cite{Shi1998} via a set of quasiparticles whose creation and annihilation operators $\hat{b}_j^{\dagger}$ and $\hat{b}_j$ are related to the particle ones via two representations of the excited-particle field operator, $\hat{\psi}_{\text{ex}}({\bf r}) = \sum_{k\neq0} \psi_k({\bf r}) \hat{a}_k = \sum_j \big( u_j({\bf r}) \hat{b}_j + v_j^*({\bf r}) \hat{b}_j^{\dagger} \big)$, and a symplectic matrix $R$ of Bogoliubov transformation
\begin{equation}    \label{BdG_transform}
    V_{\hat{a}} = R \ V_{\hat{b}},
    \ 
    V_{\hat{a}} \equiv (.., \hat{a}_k^{\dagger}, \hat{a}_k, ..)^T,
    \
    V_{\hat{b}} \equiv (.., \hat{b}_j^{\dagger}, \hat{b}_j,..)^T.
\end{equation}
The superscript $T$ stands for a transpose operation. 
The vectors $V_{\hat{a}}$ and $V_{\hat{b}}$ consist of the creation and annihilation operators of the particles and quasiparticles, respectively. 

The condensate obeys the Gross-Pitaevskii equation 
\begin{equation}
\hat{\mathscr{L}} \psi_0 = 0; \quad \hat{\mathscr{L}} \equiv -\frac{\hbar ^2 \Delta}{2M} + U + g \la N_0 \ra \psi_0^2 + 2g \nex -\mu .
    \label{L}
\end{equation}
Here $\Delta$ is the three-dimensional Laplace operator, $g=4\pi \hbar^2 a/m$ an interaction constant, $m$ a particle mass, $\mu$ a chemical potential, $\la N_0 \ra$ a mean number of particles in the condensate, and $ \nex ({\bf r}) = \la \hat{\psi}^{\dagger}_{\text{ex}}({\bf r}) \hat{\psi}_{\text{ex}}({\bf r}) \ra$ is a mean density profile of the excited particle fraction.
For simplicity of formulas, here we set $\psi_0$ to be real-valued. 
The angles stand for a statistical averaging, $\la \ldots \ra = \text{Tr} \{ \ldots \hat{\rho} \}$.    
The quasiparticle wave function $\{ u_j,v_j \}$ of an energy $E_j$ is the solution of the Bogoliubov-de Gennes equations:
\begin{equation}    \label{GP+BdG}
\begin{split}
    &\hat{\mathscr{L}} u_j + g \la N_0 \ra \psi_0^2({\bf r}) (u_j+v_j) = + E_j u_j,   
    \\
    &\hat{\mathscr{L}} v_j + g \la N_0 \ra \psi_0^2({\bf r}) (u_j+v_j) = - E_j v_j .
\end{split}
\end{equation}
The wave functions are normalized to unity as follows: $\int_V |\psi_0|^2 d^3{\bf r} = 1$, $\int_V \left( |u_j|^2-|v_j|^2 \right) d^3{\bf r} = 1$.

We assume that the excited states are orthogonal to the condensate. 
This can be gained via an ad hoc orthogonalization procedure \cite{Hadzibabic2019excitations}. 
There is a more convenient choice of such states as the solutions of a single-particle BEC-modified Schr$\ddot{\text{o}}$dinger equation \cite{HZF,PRA2020}, $\hat{\mathscr{L}} f_k =\epsilon_k f_k$, in which the potential is modified by the condensate (obviously, $f_0 = \psi_0$). 
The set $\{ f_k({\bf r})| k=1,2,... \}$ forms a complete basis in the single-particle Hilbert space.

This mean-field approach accounts for interactions and is not reduced to just a modification of the excitation spectrum. Via nonlinear Gross-Pitaevskii and Bogoliubov-de Gennes equations, the bare particles (atoms) acquire Bogoliubov couplings and form the quasiparticles - superpositions of many bare particles. The eigenvectors (quasiparticles) are no less important than their eigenvalues (excited energies), especially, since in the experiments the detectors count the real atoms (bare particles), not the virtual energy eigenvectors (quasiparticles). This fact brings into the game an interplay between the interference and interactions of bare particles. This interplay is the ultimate cause for (i) a self-generation of the squeezed states by a quantum many-body interacting system even in the thermal state and (ii) an appearance of quantum advantage in atomic boson sampling revealed in this Letter.

Consider occupations of any basis particle states $\{ \psi_k| k\ne 0 \}$ in this Hilbert space. They are described by the Hermitian operators $\hat{n}_k = \hat{a}_k^{\dagger}\hat{a}_k$ and can be measured by the appropriate detectors projecting particles onto these states. 
We calculate the joint probability distribution of these observables $\{ \hat{n}_k| k\ne 0 \}$ as follows
\begin{equation}	\label{CF_PDF} 
\begin{split}
&\rho(\{ n_k \}) = \int_{-\pi}^{\pi}...\int_{-\pi}^{\pi} e^{-i \sum_k u_k n_k} \Theta (\{ u_k \}) \prod_k \frac{du_k}{2\pi} ,
\\
    &\Theta(\{ u_k \}) =    \la e^{i\sum_k u_k\hat{n}_k} \ra
                    \equiv \text{Tr} \big\{ e^ {i\sum_k u_k\hat{n}_k} \hat{\rho} \big\}. 
\end{split}
\end{equation}
Utilizing the method employed in \cite{PRA2020} but assigning now an individual argument $z_k=e^{iu_k}$ to each excited state, we get \cite{SM} the characteristic function of this distribution:
\begin{equation}
    \Theta = \frac{1}{\sqrt{\det [1 - (Z-1)G]}}, 
    \; \;
    G_{r,k}^{r',k'} = \langle \ : \hat{a}_{r,k}^{\dagger} \hat{a}_{r',k'} : \ \rangle . 
\label{CF}
\end{equation}
$G$ is the covariance matrix with entries $G_K^{K'}$, enumerated by double indices $K = (r,k)$ for rows and $K' =(r',k')$ for columns and equal to normally-ordered (note colons) averages of a product of two creation/annihilation operators.
Nambu-type index $r$ acquires two values: 1, 2. For any operator $\hat{\mathscr{O}}$, it denotes that same operator, $\hat{\mathscr{O}}_r = \hat{\mathscr{O}}$, if $r=1$ or its Hermitian conjugate, $\hat{\mathscr{O}}_r = \hat{\mathscr{O}}^{\dagger}$, if $r=2$. It is related to the $(2\times 2)$-block structure of the matrix.

The variables form a diagonal matrix $Z = \textrm{diag} (\{ z_K \})$ which contains pairs of the same variable $z_{r,k}=z_k=e^{iu_k}$ along the diagonal and has a size that is twice the number $M$ of excited particle states in the considered miniband. 

The result (\ref{CF}) is truly general and universal: It is valid for the joint occupations of any number of states $M$ by any number of the interacting Bose particles $N-\la N_0 \ra$. Its derivation via quasiparticles \cite{SM} involves the covariance matrix expressed via the unitary Bogoliubov matrix $R$,
\begin{equation}
   G =  R \big(D + \frac{1}{2}\big) A R^T A - \frac{1}{2}; A = \bigoplus_j \sigma_x, 
   D = \bigoplus_j \frac{\sigma_0}{e^{E_j/T}-1}.
   \label{CM}
\end{equation}
The block-diagonal matrices $A$ and $D$ hold the Pauli matrix $\sigma_x$ and identity $(2\times2)$-matrix $\sigma_0$, respectively.

We derived Eq.~(\ref{CF}) also within the microscopic theory of critical phenomena \cite{PLA2015,PhysScr2015,arXivIsing2016} via the method of the recurrence equations for the partial operator contractions, unrelated to the Bogoliubov-Popov picture of the BEC-condensed gas. 
It is valid for any system of the interacting unconstraint bosons in an equilibrium state described by any normally-ordered covariance matrix $G$, that is, for any state $\hat{\rho} = e^{-\hat{H}/T}/{\rm Tr}\{ e^{-\hat{H}/T} \}$. 
The joint occupation probability distribution is given by the mixed derivatives
\begin{equation}
\rho(\{ n_k \}) = \prod_k \frac{ \partial^{n_k}}{n_k!\partial z_k^{n_k}} \Theta \Big|_{\{ z_k=0 \}}, \ \{z_k \equiv e^{iu_k}|k=1,2,... \}.
\label{pdf}
\end{equation}

Here we point up that the result for the characteristic function (\ref{CF}) and its cumulant analysis provide the most efficient, canonical method for characterizing such a complex joint distribution and distinguishing it from various mockups via generating cumulants \cite{PRA2000} $\{ \tilde{\kappa}_{\{ m_k \}}|m_k = 1,2,...\}$ defined by the Taylor expansion 
\begin{equation}
\ln \Theta = \sum_{\{ m_k \}} \tilde{\kappa}_{\{ m_k \}} \prod_k \frac{(e^{iu_k}-1)^{m_k}}{m_k!}
    \label{cumulants-def}
\end{equation}
and directly related to the moments and cumulants of the distribution. 
Let us use the Permanent Master Theorem,  
\begin{equation}
\frac{1}{\det (I-ZC)} = \sum_{\{ s_K \}} \Big[ (\textrm{per} \ C(\{ s_K \})) \prod_K \frac{z_K^{s_K}}{s_K!} \Big], 
    \label{MMT}
\end{equation}
of MacMahon \cite{MacMahon1916,Percus1971}. A double index $K=(r,k)$ runs over all rows of a matrix $C(\{ s_K \})$, $\{ s_K \}$ is a set of non-negative integers. It is valid for any, even not pair-wise equal variables $z_{1,k}$, $z_{2,k}$. 
The coefficients of this Taylor expansion are given by the permanent of the $C(\{ s_K \})$ which is the $C$ with the $K$-th row and $K$-th column replaced by the same $K$-th row and $K$-th column $s_K$ times.

If we had $2M$ stochastic variables, i.e., the number $2M$ of independent variables $z_K$ in the matrix $Z$ was equal to the number of matrix rows, and the square root in Eq.~(\ref{CF}) for the characteristic function was absent, then we would at once conclude that the occupation probability,
\begin{equation}
\rho'(\{ n_K \}) = \frac{\textrm{per} \ (C(\{ s_K \}))}{\det (1+G) \prod_K s_K!}, \quad C=AG(1+G)^{-1},
 \label{rho-aux}
\end{equation}
is given by a permanent of the extended matrix $C(\{ s_K \})$ built of the matrix $C$ as stated above. The characteristic function $\Theta'$ of such an auxiliary probability distribution lays an extended set of the generating cumulants $\tilde{\kappa}'_{\{ m_K \}}$,
\begin{equation}
\ln \Theta' = \sum_{\{ m_{r,k} \}} \tilde{\kappa}'_{\{ m_{r,k} \}} \prod_{r,k} \frac{(e^{iu_{r,k}}-1)^{m_{r,k}}}{m_{r,k}!}.
    \label{aux-cumulants-def}
\end{equation}
Since in Eq.~(\ref{CF}) (a) there are two times less independent variables because $z_{1,k}=z_{2,k}=z_k$ and (b) the square root adds a prefactor $1/2$ for $\ln \Theta$, we get the true generating cumulants as the simple finite sums of the auxiliary ones:
\begin{equation}
\tilde{\kappa}_{\{m_k \}} = \frac{1}{2} \sum_k \sum_{m_{1,k}=0}^{m_k} \Big[ \tilde{\kappa}'_{\{ m_{1,k}, m_k -m_{1,k} \}} \prod_{k'} \binom{m_{1,k'}}{m_{k'}} \Big] .
    \label{cumulants}
\end{equation}
Here a pair of the arguments $m_{r,k}, r=1,2$, in $\tilde{\kappa}'_{\{ m_{r,k} \}}$ is written explicitly for the case when $m_{1,k} +m_{2,k} =m_k$; $\binom{m_{1,k}}{m_k} = m_k!/(m_{1,k}!m_{2,k}!)$ is a binomial coefficient.

It is immediate to get the distribution~(\ref{pdf}) explicitly as
\begin{equation}
\rho(\{ n_k \}) = \frac{\textrm{haf} \ (\tilde{C}(\{ n_k \}))}{\sqrt{\det (1+G)} \prod_k n_k!}, \ C=AG(1+G)^{-1},
    \label{pdf=hafnian}
\end{equation}
from Eq.~(\ref{CF}) via the Wick's theorem which is well known in the quantum field theory \cite{Wick1950,FetterWalecka} and is equivalent, in this case, to the Hafnian Master Theorem \cite{SM} 
\begin{equation} \label{HMT}
\frac{1}{\sqrt{\det (1+(1-Z)G)}} = \sum_{\{ n_k \}} \frac{\textrm{haf} \ (\tilde{C}(\{ n_k \}))}{\sqrt{\det (1+G)}} \prod_{k} \frac{z_k^{n_k}}{n_k!}.
\end{equation}
In fact, the hafnian \cite{Barvinok2016,Mansour2015,footnoteHaf} was introduced in \cite{Caianiello1953,Caianiello1973} as a notation for a Wick's sum of all possible products of $n$ two-operator contractions (averages) in a given product of $2n$ creation/annihilation operators. 
Here the hafnian is a function of the $(2n \times 2n)$-matrix $\tilde{C}(\{ n_k \})$, $n = \sum_k n_k$, built of the matrix $C$, Eq.~(\ref{pdf=hafnian}), via replacing the $k$-th pair of rows and the $k$-th pair of columns by $n_k$ pairs of the same $k$-th pair of rows and by $n_k$ pairs of the same $k$-th pair of columns, respectively. 
The MacMahon Master Theorem (\ref{MMT}) follows from (\ref{HMT}) as a particular case.

The distribution (\ref{pdf=hafnian}) was also derived via a standard phase-space method \cite{WignerReviewScully,Barnett1996} and applied to the photon sampling of Gaussian states in \cite{Hamilton2017,Hamilton2019}. 
The phase-space method had been applied in BEC statistics in \cite{Englert2002} for rederiving an original result of \cite{PRA2000} on the statistics of a Gaussian state of the atomic modes squeezed by Bogoliubov coupling. 
In \cite{SM}, we use the method of \cite{Englert2002}.    

Computing the cumulants and joint probability distribution for the excited particle occupations is a $\#P$-hard problem due to a $\#P$-hard complexity of computing the permanents \cite{Valiant1979}, while the $(2\times 2)$-block structure of the matrices $A,D,G,C$ and the presence of the square root in Eq.~(\ref{CF}) (the prefactor 1/2 in Eq.~(\ref{cumulants})) just modify it a bit to a similar, hafnian $\#P$-hard complexity, Eq.~(\ref{pdf=hafnian}).

For proving quantum advantage, a $\#P$-hardness persisting for the average case is needed as well as an analysis of the approximate case is required. Fortunately, such analyses for the atomic and photonic boson samplings are very similar since the universal result in Eqs. (\ref{CF}), (\ref{pdf=hafnian}) put these two samplings on the same footing, both with respect to expressing the joint probability via the hafnian and ranging complex-valued matrices associated with the sampling. For the interferometer, a wide range of complex unitary matrices appears due to varying its partial modes via adjusting phase shifts and couplings. For the BEC, a wide range of complex matrices appears due to varying the partial atomic wave functions (excited states) assigned to be projected upon for detectors measuring their occupations. In particular, the "hiding" technique employed in the proof of quantum advantage works equally well for both samplings. Besides, in both samplings, the squeezing parameters of the matrix under the hafnian are controllable via adjusting squeezing in the input sources in the optical interferometer or the condensate wave function and Bogoliubov couplings by changing the trapping potential, interaction (via Feshbach resonances \cite{Chin2010}), temperature or number of trapped atoms. We skip repeating such $\#P$-hardness analyses, see \cite{Aaronson2013,Shi2021,Shchesnovich2019,Hamilton2019,Chin2018,Quesada2018,Huh2019,Villalonga2021,BentivegnaBayesianTest2015,Renema2018,Renema2020,Popova2021,Qi2020}.

Remarkably, the result (\ref{CF}) for the characteristic function is universal in the sense that it has the same form for any marginal, restricted subset of the excited particle states. 
Averaging over the rest excited-state occupations is achieved by setting all irrelevant variables $z_{k'}$ equal to zero and keeping just those rows and columns in the matrices $A, D, G, C$ which are associated with the chosen marginal subset of excited states. The $\sharp$P-hardness disappears if the $C$ is degenerate, e.g., there is no interaction or the condensate is uniform \cite{SM,footnoteCE,PRA2010,Entropy2021}.    

{\it Testing boson sampling in the atomic BEC trap and comparing it with photonic-interferometer experiments.} -- The atomic BEC trap can be viewed as a boson-sampling platform alternative to a photonic interferometer. In both systems, the output multivariate statistics is $\sharp$P-hard to compute and is associated with the hafnians of complex-valued, easily controllable matrices. This allows one to vary the output statistics over a wide range.  

The excited atoms naturally fluctuate and are squeezed inside the trap even in the thermal state. This allows one to eliminate the nonequilibrium state/dynamics and sophisticated external sources of squeezed or single bosons (which were required for photonic sampling) from the atomic sampling experiments \cite{SM}. So, the losses of bosons on the input-output propagation, which constitute the main limitation factor in photonic sampling, are no more an issue for atomic sampling. It remains just to measure the distribution of atoms over the excited state subset by means of appropriate detectors.

It would be very interesting to study experimentally various phenomena associated with boson sampling by simultaneously measuring excited state occupations, say, via a multi-detector imaging based on the light transmission through or scattering from the atomic cloud \cite{footnoteImaging}. The transmission imaging is based on the absorption or dispersion caused by atoms \cite{Jacqmin2010,Kristensen2019,Kristensen2017,Okeke2021,AspectDensityFluctPRL2006}. A scattering or fluorescence imaging \cite{RaizenBECstatisticsPRL2005}, including Raman one, could be facilitated by exciting modes, mimicking excited states, via lasers, cavities. Such experiments could be devised similar to optical imaging of the local atom-number fluctuations \cite{Pit2011,Castin,Jacqmin2010,Armijo2010,AspectDensityFluctPRL2006,RaizenBECstatisticsPRL2005,Dotsenko2005,Schlosser2002}.  

Measuring with a single atom resolution is challenging, but a nearly single atom resolution had been achieved \cite{RaizenBECstatisticsPRL2005,Dotsenko2005,Schlosser2002,Raizen2009}. Though, it is not required for showing quantum advantage since boson sampling is $\#P$-hard for computing even if it is done with threshold detectors. Such detectors provide just two measurement outcomes – either zero or non-zero occupation in a given mode. The threshold boson sampling is described by torontonians (their computing is not easier than computing the hafnians) and still possesses quantum advantage \cite{Quesada2018,Shi2021,Villalonga2021}.

{\it Conclusions.} -- (i) We found the characteristic function (\ref{CF}) for the fluctuations of the excited-particle-state occupations. It is the universal determinantal function which is easy to compute in polynomial time. 

(ii) We found the Hafnian Master Theorem (\ref{HMT}) which is a hafnian's analog and generalization of the famous MacMahon Master Theorem on the matrix permanents.

(iii) Computing a Fourier transform of the characteristic function, that is the corresponding joint probability distribution, amounts to computing the permanents and hafnians \cite{SM,Rudolf2009} and is $\sharp$P-hard. The latter implies a quantum advantage of the many-body BEC fluctuations. Clearly, the $\sharp$P-hardness is due to multiple Fourier integration (cf. a permanent's integral representation \cite{Entropy2020}). 

(iv) Conceptually, the particle sampling in the excited states of a BEC trap and the Gaussian, squeezed photon sampling in an interferometer are on the same footing.    

(v) There is a remarkable difference between the two: Due to many-body fluctuations and interparticle interaction, the particle sampling in the BEC trap possesses the quantum advantage even in a thermal, equilibrium state (without any particle source) while a nonthermal photon source is required in the linear interferometer.

(vi) It is worth to employ the characteristic function and cumulant analysis, which constitute a well-known comprehensive tool in statistics and are sketched above for the boson sampling, for (a) ruling out mockups, such as with non-squeezed states or distinguishable bosons, (b) verifying that incoherent processes, boson loss, technical noise, detector dark counts, other imperfections do not wash out the $\sharp$P-hardness of sampling \cite{footnoteMockups,BentivegnaBayesianTest2015,Renema2018,Renema2020,Popova2021}.

(vii) Especially promising are boson-sampling experiments with the multi-qubit BEC trap formed by a finite number $Q$ of qubit wells. The results (\ref{CF}), (\ref{cumulants}), (\ref{pdf=hafnian}) show that the many-body statistics of the excited atom occupations in the BEC trap offers a quantum simulation of the $\sharp$P-hard problem of boson sampling on the platform alternative to the photonic interferometer platform \cite{SM}.

Overall, the analysis above goes far beyond the existing photon sampling studies in a linear interferometer. It allows researchers from different fields to initiate exploring/designing the $\sharp$P-hard complexity in their own interacting systems of various particles and fields.

We acknowledge the support by the Russian Science Foundation (grant 21--12--00409).

\clearpage
\onecolumngrid
\section*{SUPPLEMENTAL MATERIAL}

\renewcommand\theequation{S-\arabic{equation}}
Here we derive the result for the characteristic function (Eq.~(5) of the main text of the Letter)
\begin{equation}
    \Theta\big(\{z_k=e^{i u_k}\}\big) = \frac{1}{\ \sqrt{\ \det [1 - (Z-1)G]\ } \ },
    \quad
    G \equiv R \Big(D + \frac{1}{2}\Big) A R^T A - \frac{1}{2}, 
    \quad 
    Z \equiv \bigoplus_{k=1}^{M} \left[  \begin{matrix}  z_k & 0 \\  
                            0 & z_k   \end{matrix}    \right].
\label{sm-CF}
\end{equation}
The symmetric block-diagonal matrices $A$, $D$ include the Pauli $(2\times 2)$-matrices $\sigma_x$, $\sigma_0$ and quasiparticle energies $E_j$:
\begin{equation} \label{sm-AD}
    A   =   \left[  \begin{matrix}  
                \sigma_x & 0 & \ldots   \\  
                0 & \sigma_x & \ldots   \\
                \ldots & \ldots & \ldots
            \end{matrix} \right] \equiv \bigoplus_{j=1}^M \sigma_x,
    \quad
    \sigma_x = \left[  \begin{matrix}  0 & 1 \\  
                                       1 & 0   \end{matrix}    \right];
    \qquad
    D = \bigoplus_{j=1}^{M} \frac{\sigma_0}{e^{E_j/T}-1},
    \quad
    \sigma_0 = \left[  \begin{matrix}  1 & 0 \\  
                                       0 & 1   \end{matrix}    \right].
\end{equation}
The matrix $R$ describes the Bogoliubov transformation from the vector $V_{\hat{b}} \equiv (.., \hat{b}_j^{\dagger}, \hat{b}_j,..)^T$ of the quasiparticle creation/annihilation operators to the vector $V_{\hat{a}} \equiv (.., \hat{a}_k^{\dagger}, \hat{a}_k, ..)^T$ of the particle creation/annihilation operators:
\begin{equation}    \label{sm-BdG_transform}
    V_{\hat{a}} = R \ V_{\hat{b}}.
\end{equation}
Since the Bogoliubov transformation preserves the Bose commutation relations for the creation/annihilation operators, the matrix $R$ has the symplectic properties, that is, it obeys the following relation involving the block-diagonal matrix $\Omega$ formed by the Pauli $(2\times 2)$-matrix $\sigma_y$:
\begin{equation}    \label{sm-RYR^T=Y}
   R \ \Omega \ R^T = \Omega,  \qquad \Omega =  \left[  \begin{matrix}  
                i\sigma_y & 0 & \ldots   \\  
                0 & i\sigma_y & \ldots   \\
                \ldots & \ldots & \ldots
            \end{matrix} \right] \equiv \bigoplus_{j=1}^M i\sigma_y, \qquad 
            i\sigma_y = \left[  \begin{matrix}  0 & +1 \\  
                                       -1 & 0   \end{matrix}    \right].
\end{equation}

Next, we prove that the matrix $G = \big(G_K^{K'}\big)$ in Eq.~ (\ref{sm-CF}) is the covariance matrix defined as the statistical average of the normally-ordered product of two particle creation/annihilation operators,
\begin{equation} \label{sm-G}
    \big( G_{r,k}^{r',k'} \big) = \big( \langle \, : \! \hat{a}_{r,k}^{\dagger} \hat{a}_{r',k'} \! : \, \rangle \big) = \begin{bmatrix}
        \ddots & \vdots & \vdots &   \\
        \cdots & \langle \hat{a}_k^{\dagger} \hat{a}_{k'} \rangle & 
                \langle \hat{a}_k^{\dagger} \hat{a}_{k'}^{\dagger} \rangle & \cdots \\
        \cdots & \langle \hat{a}_k \hat{a}_{k'} \rangle & 
                 \langle \hat{a}_{k'}^{\dagger} \hat{a}_k \rangle & \cdots \\
        & \vdots & \vdots & \ddots
\end{bmatrix}, \qquad G_{r,k}^{r',k'} = \Big(G_{r',k}^{r,k'}\Big)^* .
\end{equation}
Its entries $G_K^{K'}$ are enumerated by the double indices $K = (r,k)$ for rows and $K' =(r',k')$ for columns. A Nambu-type index $r$ (or $r'$) acquires two values: 1, 2. For any operator $\hat{\mathscr{O}}$, it denotes that same operator, $\hat{\mathscr{O}}_r = \hat{\mathscr{O}}$, if $r=1$ or its Hermitian conjugate, $\hat{\mathscr{O}}_r = \hat{\mathscr{O}}^{\dagger}$, if $r=2$. It is related to the $(2\times 2)$-block structure of the matrix. We assume $\la \hat{a}_k \ra =0$. The diagonal matrix $Z = \textrm{diag} (\{ z_K \})$ consists of the pairs of the same variable $z_{r,k}=z_k=e^{iu_k}$ along the diagonal. 

We use the notations of the main text of the Letter and mostly consider the system of a finite number, $M$, of the excited particle modes. So, the $A,D,R,G,\Omega$, and $Z$ are essentially the $(2M \times 2M)$-matrices. However, the method below can be easily extended to the case of an arbitrary countable set of an infinite number of excited modes.

In the section II, we derive the Hafnian Master Theorem.

In the section III, we discuss the $\sharp$P-hardness of the atomic boson sampling. 

\section*{S-I. The characteristic function of the joint probability distribution \\of the excited-state particle numbers}

{\bf Calculation of the characteristic function} is similar to the one described in \cite{PRA2020} and is based on the Wigner transform technique \cite{Englert2002,WignerReviewScully,Barnett1996}. The Wigner transformation casts an operator-valued function $F(\hat{a}^\dagger,\hat{a})$ of the creation and annihilation operators $\hat{a}^\dagger$ and $\hat{a}$ into a complex-valued function $W_F$ of the associated variables $\alpha^*$ and $\alpha$ as follows
\begin{equation}
    W_F(\alpha^*,\alpha) = \int_{\mathbb{C}} e^{-\gamma \alpha^* + \gamma^* \alpha} \
        \text{Tr}   \left( 
                e^{\gamma \hat{a}^\dagger - \gamma^* \hat{a}}     
                    F(\hat{a}^\dagger,\hat{a}) 
                \right)
        \frac{d^2 \gamma}{\pi}.
\end{equation}
It allows one to represent the trace of an operator product $\hat{F} \hat{G}$ via a complex integral, $\text{Tr}\,(\hat{F} \, \hat{G}) = \pi^{-1}\int W_F \, W_G \ d^2 \alpha$. The above formulas are written in the single-mode case. In the multi-mode case, they include the multiple integrals. In particular, the characteristic function, $\Theta(\{ u_k \}) \equiv \text{Tr} \big(e^ {i\sum_k u_k\hat{n}_k} \hat{\rho} \big)$, has the following Wigner representation
\begin{equation}  \label{sm-CF=WW}
    \Theta\big(\{u_k\}\big) = \int_{\mathbb{C}^M} W_{\{n_k\}}\left(\{\alpha^*_k, \alpha_k\}\right)
                                W_{\rho}\left(\{\alpha^*_k, \alpha_k\}\right)
                        \prod_{k=1}^M \frac{d^2\alpha_k}{\pi}.
\end{equation}
It is easy to calculate the Wigner transform of the statistical operator $\hat{\rho} = e^{-\sum_j E_j \hat{b}^\dagger_j \hat{b}_j/T}/\text{Tr}\, \{ e^{-\sum_j E_j \hat{b}^\dagger_j \hat{b}_j/T} \}$ as follows
\begin{equation}    \label{sm-Wr}
\begin{split}
    W_{\rho}\big(\{\beta^*_j, \beta_j\}\big) &= \prod_{j=1}^M \left( 2 \tanh\frac{E_j}{2T} \right)
        \exp {\left( -2 \beta^*_j \beta_j \tanh \tfrac{E_j}{2T}  \right)}
    = e^{-V_{\beta}^T B V_{\beta}} \prod_{j=1}^M \left( 2 \tanh\frac{E_j}{2T} \right); 
    \\
    V_{\beta} &\equiv (\ldots, \beta_j^*, \beta_j, \ldots)^T,
    \qquad
    B = \bigoplus_{j=1}^M \: \sigma_x \tanh\tfrac{E_j}{2T}.
\end{split}    
\end{equation}
Here the complex variables $\beta_j^*$ and $\beta_j$ are associated with the quasiparticle operators $\hat{b}_j^\dagger$ and $\hat{b}_j$, respectively, and constitute the vector $V_{\beta}$ of the size $2M$ which is the counterpart of the vector $V_{\hat{b}}$ introduced in Eq.~(\ref{sm-BdG_transform}) above.

The Wigner transform of the operator $\exp \big( i \sum_k u_k \hat{a}^\dagger_k \hat{a}_k \big)$, whose average equals the characteristic function, is
\begin{equation}    \label{sm-WN}
\begin{split}
    W_{\{n_k\}}\big(\{\alpha^*_k, \alpha_k\}\big) &= \prod_{k=1}^M \frac{2}{z_k + 1} 
                    \exp{ \left(2 \alpha^*_k \alpha_k \frac{z_k-1}{z_k+1} \right)} 
                =  \exp \left(V_{\alpha}^T \frac{Z-1}{Z+1} A V_{\alpha} \right) \prod_{k=1}^M \frac{2}{z_k+1}; 
    \\
    V_{\alpha} &\equiv (\ldots, \alpha_k^*, \alpha_k, \ldots)^T,
    \qquad 
    \frac{Z-1}{Z+1} = \bigoplus_{k=1}^M \: \sigma_0 \: \frac{z_k-1}{z_k+1} = \bigoplus_{k=1}^M \: \sigma_0 \: \frac{e^{iu_k}-1}{e^{iu_k}+1}.
\end{split}            
\end{equation}
Similar to Eq.~(\ref{sm-Wr}), the complex variables $\alpha_k^*$ and $\alpha_k$ are associated with the particle operators $\hat{a}_k^\dagger$ and $\hat{a}_k$, respectively, and constitute the vector $V_{\alpha}$ of the size $2M$ which is the counterpart of the vector $V_{\hat{a}}$ introduced in Eq.~(\ref{sm-BdG_transform}) above. Each argument of the characteristic function $u_k$, $k=1,\ldots, M$, appears, in the form of the exponential variable $z_k = e^{iu_k}$, twice in the entries of the $k$-th $(2\times2)$-block of the block-diagonal $(2M\times 2M)$-matrix $(Z-1)(Z+1)^{-1}$.

Also, we get the Wigner transform of the auxiliary characteristic function $\Theta'$, which is the Fourier transform of the auxiliary joint probability distribution $\rho' (\{ n_K \})$ in Eq.~(10) 
of the main text of the Letter, in a similar form 
\begin{equation}    \label{sm-WN_mod}
    W'_{\{ n_k \}}\big(\{\alpha^*_k, \alpha_k\}\big) =  
                    \exp \left( V_{\alpha}^T \frac{Z'-1}{Z'+1} A V_{\alpha} \right) \prod_{k=1}^M \frac{2}{\sqrt{(z_{1,k}+1)(z_{2,k}+1)}}; \qquad Z' = \bigoplus_{k=1}^M 
                \left[  \begin{matrix}  z_{1,k}     &   0 \\  
                                        0   &   z_{2,k} \end{matrix}    \right].
\end{equation}
Here, instead of the single variable $z_k = e^{iu_k}$, we assign to each mode $k$ a pair of independent variables $z_{1,k} \equiv e^{i u_{1,k}}$ and $z_{2,k} \equiv e^{i u_{2,k}}$ denoted by means of the Nambu-type index $r = 1, 2$.

Now we employ the property of the Wigner transform highlighted in \cite{Englert2002}: The linear similarity transformation of the operator functions carries over to their Wigner functions. It allows us to find the Wigner transform $W_{\rho}\left(\{\alpha^*_k, \alpha_k\}\right)$ for Eq.~(\ref{sm-CF=WW}) by substituting variables $V_{\beta} = R^{-1} V_{\alpha}$ into Eq.~(\ref{sm-Wr}). As a result, Eq.~(\ref{sm-CF=WW}) takes the following explicit form
\begin{equation}
    \Theta\big(\{z_k\}\big) = \int_{\mathbb{C}^M} \exp \left[ - V_\alpha^T (\tilde{B} - \frac{Z-1}{Z+1} A) V_\alpha \right] \prod_{k=1}^M \frac{4 \tanh(E_k/2T)}{z_k+1} \frac{d^2\alpha_k}{\pi}, \qquad \tilde{B} \equiv (R^T)^{-1} B R^{-1} .
\end{equation}
The matrix $\tilde{B}$ is the matrix $B$ written in the "particle" basis as opposed to the "quasiparticle" basis.

Changing the integration variables to $\text{Re} \, \alpha_k$ and $\text{Im} \, \alpha_k$ and applying a well-known formula for the Gaussian integral $\int_{\mathbb{R}^n} \exp\big(-{\bf x}^T S {\bf x}\big) \prod_{j=1}^n dx_j = \pi^{n/2}\big/\sqrt{\det S}$ with a symmetric matrix $S = S^T$ whose real part $\text{Re}\,S$ is positively definite, we get
\begin{equation} \label{sm-GI}
    \Theta  =   \frac{
                            2^M \prod_{j=1}^M \tanh \frac{E_j}{2T}
                        }{
                            \sqrt{(-1)^M \det (\tilde{B} - \frac{Z-1}{Z+1} A) \prod_{k=1}^M (z_k+1)^2}
                        }                
            =   \frac{  2^M }{
                            \sqrt{
                                (\det \tilde{B})^{-1} \det \big(\tilde{B} - \frac{Z-1}{Z+1} A \big) \det (Z+1)}
                        }.                          
\end{equation}
The last equality follows from representing the left-hand-side products as the determinants of the appropriate matrices:
\begin{equation}
    \prod_{j=1}^M \tanh^2 (E_j/2T) = (-1)^M \det B = (-1)^M \det \tilde{B}, 
    \qquad
    \prod_{k=1}^{M} \left(z_k + 1\right)^2 = \det (Z+1) .
\end{equation}
The matrices $B$ and $\tilde{B}$ have equal determinants, $\det B = \det \tilde{B}$, since the Bogoliubov transformation preserves the commutation relations and, hence, its matrix $R$ is simplectic that implies $\det R = \det R^T = 1$.
Multiplying the matrices in the denominator, we get
\begin{equation} \label{sm-CFzk}
    \Theta \big(\{z_k\}\big)    =   \frac{1}
                            {\sqrt{                           
                            \det 
                                \left( 
                                    \frac{\tilde{B}^{-1} A + 1}{2}
                                    -
                                    Z \frac{\tilde{B}^{-1} A - 1}{2}
                                \right)
                                }
                            } = \frac{1}
                            {\sqrt{\det 
                                \left( 
                                    1 - (Z-1)\frac{R B^{-1} R^T A - 1}{2}
                                \right)
                                }
                            }.
\end{equation}
The inverse of the block-diagonal matrix $B$ is straightforward to calculate as $B^{-1} = A (1 + 2D)$.
As a result, Eq.~(\ref{sm-CFzk}) acquires the form of Eq.~(\ref{sm-CF}). This completes the proof of the first part of Eq.~(\ref{sm-CF}). 

The formula for the characteristic function (\ref{sm-CF}) is derived above for the case of a finite number of excited states $M$. However, the final result does not explicitly depend on the dimension $M$ of the Hilbert space on which the bosons live. So, the formula in Eq.~(\ref{sm-CF}) can be also applied to a Bose system with an infinite number of the excited states. Of course, the finite-size matrix definitions and the finite products employed above should be modified accordingly in order to fit the case of an infinite countable dimension.

The formula for the characteristic function $\Theta'(\{z_{1,k},z_{2,k}\})$ of the auxiliary joint probability distribution $\rho' (\{ n_K \})$,
\begin{equation}
    \Theta'(\{z_{1,k},z_{2,k}\}) =   \frac{1}
                            {\sqrt{\det
                                \left(1 - (Z'-1) G \right)
                                }
                            }, 
    \qquad
    Z' \equiv \bigoplus_{k=1}^{M} \left[ \begin{matrix}  z_{1,k} &   0   \\  
                                                        0       &   z_{2,k}
                                        \end{matrix} \right],
\end{equation}
in Eq.~(10) 
of the main text of the Letter has been derived similarly. One just need to use the Wigner transform $W'_{\{ n_k \}}$, Eq.~(\ref{sm-WN_mod}), instead of the $W_{\{ n_k \}}$, Eq.~(\ref{sm-WN}). Also, there are now two different variables $z_{1,k} \equiv e^{i u_{1,k}}$ and $z_{2,k} \equiv e^{i u_{2,k}}$, which are the entries of the diagonal matrix  $Z'$. The latter replaces the matrix $Z$ in Eq.~(\ref{sm-CF}).
\\

{\bf Derivation of the formula for the normally-ordered covariance matrix $G$} (defined in Eq.~(\ref{sm-G})), that is, the second part of Eq.~(\ref{sm-CF}), can be done via the auxiliary $(2M\times 2M)$-matrices
\begin{equation} \label{sm-MM-def}
    M_{\hat{a}} \equiv V_{\hat{a}} \ V_{\hat{a}}^T = 
    \begin{bmatrix}
        \ddots & \vdots & \vdots &   \\
        \cdots & \hat{a}_k^{\dagger} \hat{a}_{k'}^{\dagger}& 
                \hat{a}_k^{\dagger} \hat{a}_{k'}& \cdots \\
        \cdots & \hat{a}_k \hat{a}_{k'}^{\dagger} & 
                \hat{a}_k \hat{a}_{k'} & \cdots \\
        & \vdots & \vdots & \ddots
    \end{bmatrix},
    \qquad  
    M_{\hat{b}} \equiv V_{\hat{b}} \ V_{\hat{b}}^T =
    \begin{bmatrix}
        \ddots & \vdots & \vdots &   \\
        \cdots & \hat{b}_j^{\dagger} \hat{b}_{j'}^{\dagger}& 
                \hat{b}_j^{\dagger} \hat{b}_{j'}& \cdots \\
        \cdots & \hat{b}_j \hat{b}_{j'}^{\dagger} & 
                \hat{b}_j \hat{b}_{j'} & \cdots \\
        & \vdots & \vdots & \ddots
    \end{bmatrix} .
\end{equation}
They are related to each other via the Bogoliubov transformation as follows
\begin{equation} \label{sm-MRMR}
 M_{\hat{a}} = R \ M_{\hat{b}} R^T .
\end{equation}
Their averages $\langle M_{\hat{a}} \rangle$ and $\langle M_{\hat{b}} \rangle$ are the covariance matrices for the particles and quasiparticles, respectively. They give the corresponding normally-ordered covariance matrices (see Eq.~(\ref{sm-G})) via the matrix $A$, Eq.~(\ref{sm-AD}), as follows
\begin{equation} \label{sm-MM}
   \big( \langle \, : \! \hat{a}_{r,k}^{\dagger} \hat{a}_{r',k'} \! : \, \rangle \big) = \big(\langle M_{\hat{a}} \rangle - Y \big) A, \qquad
   \big( \langle \, : \! \hat{b}_{r,k}^{\dagger} \hat{b}_{r',k'} \! : \; \rangle \big) = \big(\langle M_{\hat{b}} \rangle - Y \big) A, \qquad Y = \bigoplus_{j=1}^M  \left[  \begin{matrix}  0 & 0 \\                 1 & 0   \end{matrix}    \right].
\end{equation}
Here the matrix $Y$ represents the Bose commutator, $[\hat{a}_k, \hat{a}_{k'}^{\dagger}] = \delta_{k,k'}$ or $[\hat{b}_k, \hat{b}_{k'}^{\dagger}] = \delta_{k,k'}$, by which the covariance matrices in Eq.~(\ref{sm-MM-def}) differ from the normally-ordered ones.
The quasiparticles are the independent, non-interacting bosons. 
Their average occupations of the energy levels $\{ E_j \}$ in the thermal, equilibrium state, $\hat{\rho} \sim \exp{(-\sum_j E_j \hat{b}^\dagger_j \hat{b}_j}/T)$, are given by the Bose-Einstein distribution $\la \hat{b}_k^{\dagger} \hat{b}_k \ra = \big(e^{E_j/T}-1\big)^{-1}$. So, the covariance matrix of the quasiparticle operators is exactly the matrix $D$ defined in Eq.~(\ref{sm-AD}), $D = \big(\langle \, : \! \hat{b}_{r,k}^{\dagger} \hat{b}_{r',k'} \! : \; \rangle \big)$.

Eqs.~(\ref{sm-MRMR}) and (\ref{sm-MM}) immediately lead to the explicit formula for the normally-ordered covariance matrix:
\begin{equation} \label{sm-GviaRDAY}
    \big( \langle \, : \! \hat{a}_{r,k}^{\dagger} \hat{a}_{r',k'} \! : \, \rangle \big) = 
    R D A R^T A + (RYR^T - Y) A.
\end{equation}
The last relation we need is the equality $R Y R^T - Y = R Y^T R^T - Y^T$ which is equivalent to Eq.~(\ref{sm-RYR^T=Y}) expressing preservation of the Bose commutation relations under the Bogoliubov transformation since $\Omega = Y^T - Y$.
Together with the identity $A = Y^T + Y$, it allows us to symmetrize the last term of the covariance matrix as follows $R Y R^T - Y = (R A R^T - A)/2$. Plugging it into Eq.~(\ref{sm-GviaRDAY}), we get the required result for the normally-ordered covariance matrix
\begin{equation}
\big( \langle \, : \! \hat{a}_{r,k}^{\dagger} \hat{a}_{r',k'} \! : \, \rangle \big)
    \equiv G = R \Big(D + \frac{1}{2}\Big) A R^T A - \frac{1}{2} .
\end{equation}
This completes the proof of the second part of Eq.~(\ref{sm-CF}).

\section*{S-II. The hafnian master theorem}

Here we give a simple derivation of the Hafnian Master Theorem for an arbitrary covariance matrix $G$ in Eq.~(\ref{sm-G}), 
\begin{equation} \label{sm-HMT}
\frac{1}{\sqrt{\det (1+(1-Z)G)}} = \sum_{\{ n_k \}} \frac{\textrm{haf} \ (\tilde{C}(\{ n_k \}))}{\sqrt{\det (1+G)}} \prod_{k} \frac{z_k^{n_k}}{n_k!} \ , \quad C=AG(1+G)^{-1} .
\end{equation}
It establishes the Taylor series of the determinantal function $1/\sqrt{\det (1+(1-Z)G)}$ over its $M$ variables $\{ z_k | k=1,..., M \}$ at the point of origin $\{ z_k=0 \}$ and is the hafnian's analog of the Permanent Master Theorem of MacMahon \cite{MacMahon1916}. Here the $(2n \times 2n)$-matrix $\tilde{C}(\{ n_k \})$, $n = \sum_k n_k$, is built of the matrix $C$ via replacing the $k$-th pair of rows and the $k$-th pair of columns by $n_k$ pairs of the same $k$-th pair of rows and by $n_k$ pairs of the same $k$-th pair of columns, respectively.

In fact, Eq.~(\ref{sm-HMT}) is an immediate consequence of the Wick's theorem well-known in the quantum field theory \cite{Wick1950,FetterWalecka}. One just need to apply the Wick's theorem to the mixed partial derivatives of the characteristic function (\ref{sm-CF}),
\begin{equation} \label{sm-derivatives}
\prod_k \frac{\partial^{n_k}}{\partial z_k^{n_k}} \frac{1}{\sqrt{\det (1+(1-Z)G)}} \Bigg|_{\{ z_k=0 \}} = \text{Tr} \Big\{ \hat{\rho} \prod_k \Big[ z_k^{\hat{n}_k-n_k} \prod_{j=0}^{n_k-1} (\hat{n}_k-j) \Big] \Big\} \Bigg|_{\{ z_k=0 \}}; \quad \hat{\rho} = \frac{e^{-\hat{H}/T}}{{\rm Tr} \{ e^{-\hat{H}/T} \}} .
\end{equation}
If taken under the quantum-mechanical statistical average in the definition of the characteristic function $\Theta (\{ z_k \}) = \text{Tr} \{ \hat{\rho} \prod_k z_k^{\hat{n}_k} \}$, the mixed derivative can be written as above, via the products of $n_k$ shifted occupation operators $\hat{n}_k-n_k+1,...,\hat{n}_k = \hat{a}_k^{\dagger}\hat{a}_k$. For each mode $k$, in virtue of the Bose commutation relation, $[\hat{a}_k, \hat{a}_{k'}^{\dagger}] = \delta_{k,k'}$, such a product is equal to the normally-ordered product of the $n_k$ annihilation operators and $n_k$ creation operators, $\prod_{j=0}^{n_k-1} (\hat{n}_k-j) = (\hat{a}_k^{\dagger})^{n_k} \hat{a}_k^{n_k}$. It suffices to find the trace in Eq.~(\ref{sm-derivatives}) for equal variables $z_k =z = e^{iu} \to 0, \ k=1,..., M$. If the operator of the total number of excited particles $\hat{N} = \sum_k \hat{n}_k$ commuted with the Bogoliubov Hamiltonian $\hat{H}$, then we would get a usual average of a product of the creation/annihilation operators over a density matrix $\hat{\rho}_{\mu} \propto e^{-(\hat{H}-\mu \hat{N})/T}$ for a system with a related grand canonical Hamiltonian $\hat{H} - \mu \hat{N}$ and a chemical potential $\mu = iuT$. Since the $\hat{N}$ and the $\hat{H}$ do not commute, the average is a bit more involved but still can be easily calculated,
\begin{equation} \label{sm-GCEderivatives}
\prod_k \frac{\partial^{n_k}}{\partial z_k^{n_k}} \frac{1}{\sqrt{\det (1+(1-Z)G)}} \Bigg|_{\{ z_k=0 \}} = \frac{1}{\sqrt{\det (1+G)}} \text{Tr} \Big\{ \ \frac{\hat{\rho} e^{\mu (\hat{N}-n)/T}}{{\rm Tr} \{ \hat{\rho} e^{\mu (\hat{N}-n)/T} \}} \prod_k \Big[ (\hat{a}_k^{\dagger})^{n_k} \hat{a}_k^{n_k} \Big] \Big\} \Bigg|_{z=0}.
\end{equation}

According to the Wick's theorem, the average (the trace) in the right hand side of Eq.~(\ref{sm-GCEderivatives}) is equal to the sum of all possible products of $n$ two-operator contractions (averages) 
\begin{equation} \label{sm-CM}
\text{Tr} \Big\{ \ \frac{\hat{\rho} e^{\mu (\hat{N}-n)/T}}{{\rm Tr} \{ \hat{\rho} e^{\mu (\hat{N}-n)/T} \}} \, : \! \hat{a}_{r,k} \hat{a}_{r',k'} \! : \, \Big\} \Bigg|_{z=0} = C_{r,k}^{r',k'}   
\end{equation}
of a given product of $2n$ creation/annihilation operators. As a result and in virtue of the hafnian's definition \cite{Mansour2015,Barvinok2016}, originally given in the quantum field theory by Caianiello \cite{Caianiello1953,Caianiello1973}, we immediately get a concise final formula,
\begin{equation} \label{sm-Wick=haf}
\text{Tr} \Big\{ \ \frac{\hat{\rho} e^{\mu (\hat{N}-n)/T}}{{\rm Tr} \{ \hat{\rho} e^{\mu (\hat{N}-n)/T} \}} \prod_k \Big[ (\hat{a}_k^{\dagger})^{n_k} \hat{a}_k^{n_k} \Big] \, \Big\} \Bigg|_{z=0} = \textrm{haf} \ \big(\tilde{C}(\{ n_k \})\big),
\end{equation}
in terms of the hafnian as in Eq.~(\ref{sm-HMT}). Calculation of the two-operator average in Eq.~(\ref{sm-CM}) via the Wigner transforms and Gaussian integrals is a straightforward exercise similar to the calculation of the characteristic function outlined in the section I of this Supplemental Material. The result for the matrix $\big(C_{r,k}^{r',k'}\big)$ in Eq.~(\ref{sm-CM}) is $C = AG(1+G)^{-1}$. It is precisely the matrix $C$ employed in the theorem (\ref{sm-HMT}). 

The only additional, though obvious trick here is to represent the $n_k$ pairs of the $k$-mode's creation/annihilation operators in Eq.~(\ref{sm-GCEderivatives}) via the $n_k$ independent, completely degenerate (with exactly the same correlation properties) modes entering the matrix $\tilde{C}(\{ n_k \})$ in Eq.~(\ref{sm-HMT}) as the $n_k$ identical/degenerate pairs of the $k$-th rows and the $k$-th columns. 

This completes the proof of the Hafnian Master Theorem, Eq.~(\ref{sm-HMT}). The latter immediately infers Eq.~(13) 
of the main text of the Letter for the joint probability distribution.

\section*{S-III. Comments on the $\sharp$P-hardness and experimental realization \\of atomic boson sampling}

In fact, an absence of the synchronized, on-demand single-photon sources for feeding the input channels of the interferometer is the reason for a recent shift from an original proposal \cite{Aaronson2011,Aaronson2013} to a Gaussian boson sampling scheme that utilizes a two-mode squeezed (or more general, Gaussian) photon input provided by already available on-demand sources based on a parametric down-conversion \cite{LundPRL2014,Bentivegna2015,PanPRL2021}. For the BEC-trap platform, such a squeezed input is provided by nature itself due to the Bogoliubov coupling even in the box trap as had been shown in \cite{PRA2000}. So, the BEC-trap platform is closer to and should be compared with the Gaussian boson sampling.

Especially promising are boson-sampling experiments with the multi-qubit BEC trap formed by a finite number $Q$ of qubit wells (see Fig. 1 below). The results (\ref{sm-CF}) and (\ref{sm-HMT}) show that the many-body statistics of the excited atom occupations in the BEC trap offers a quantum simulation of the $\sharp$P-hard problem of boson sampling on the BEC-trap platform alternative to the photonic interferometer platform. The single-qubit case $Q=1$ corresponds to a trap with just two quasi-degenerate condensates. The case of a few qubit wells, $Q=2,3,4,...$, promises discovery of new quantum effects beyond a particle analog of the simple Hong-Ou-Mandel one and doable at the present stage of the magneto-optical trapping and detection technology. Such experiments would be tremendously valuable for understanding fundamental properties of the many-body quantum systems directly relevant to the quantum advantage. The ultimate experiments with an increasingly large number of qubits, $Q \gg 1$, addressing the $\sharp$P-hard problem are very challenging. Yet, they seem to be within reach and could hit the quantum advantage.

\begin{figure}[h]
\includegraphics*[width=15 cm]{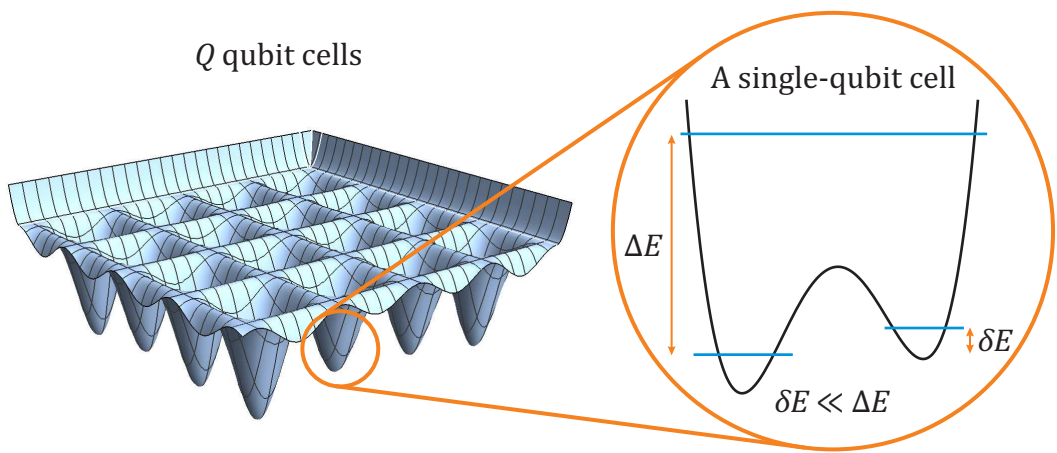}
\caption{The multi-qubit BEC trap: A sketch of the geometry of its trapping potential in the case of a two-dimensional lattice built of the $Q$ single-qubit cells. Each single-qubit cell is formed by a double-well potential featuring two close lower energy levels separated from the higher energy levels by the energy gap $\Delta E$ much larger than the lower-energy splitting $\delta E$. For clarity's sake, an inhomogeneous underlying (background) potential, designed for controlling the condensate profile and Bogoliubov couplings, as well as the high potential walls at the outer borders of the multi-qubit trap are not shown.}
\end{figure}

Detecting a particle number in each excited state can be facilitated by rising the total number $N$ of particles loaded into the trap since the excited-state occupations scale as $(N-\la N_0\ra)/M$. Rising $N-\la N_0\ra$, say, from $10^2$ to $10^4$ multiplies the occupations by 100. The asymptotic parameter of complexity is the number of Bogoliubov-coupled excited states $M$ which is similar to the number of channels in the interferometer. It is neither the total number of particles in the BEC trap $N$ nor the number of bosons (noncondensed particles) in the system $N-\la N_0\ra$. 

The experiments could be aimed at boson sampling of occupations of any-basis excited particle states, not necessary, say, single particle states of an empty trap, and even any subset of states (irrespective to the other states) or a set of groups (bunches) of states, that is, not necessary all states or each state, respectively, of the lower miniband formed by the qubit-well states. Such "incomplete" experiments on a marginal or course-grained, respectively, particle-number distribution should be the first to test the quantum advantage of the joint occupations statistics of the excited states. The related "incomplete" statistics is given by the same general formula in Eq. (\ref{sm-CF}) due to its universality. 

A reduction to computing a permanent is known also for the transition amplitude of a quantum circuit in a universal quantum computer \cite{Rudolf2009}. This fact puts the $\sharp$P-hardnesses of (i) the quantum statistics in a BEC trap and (ii) the universal quantum computer on the same footing.

{\bf When the $\sharp$P-hardness of the atomic boson sampling disappears?} -- First, if the interparticle interaction vanishes, the problem is reduced to a diagonal matrix with a trivial Bogoliubov transformation $R = 1$ that corresponds to the independent fluctuations in the occupations of the excited particle states. 
So, the aforementioned $\sharp$P-hard complexity vanishes in an ideal Bose gas within the grand canonical ensemble approximation. In the canonical ensemble, some nontrivial correlations between equilibrium occupations of the excited particle states of the trap exist even in the ideal gas due to the total particle number constraint, $N= \textrm{const}$. They are related to the known critical fluctuations in the total noncondensate or condensate occupation in the ideal gas confined in a mesoscopic trap \cite{JStatPhys2015,PRA2010}. 

Second, if the condensate is uniform, the Bogoliubov coupling reduces to just coupling inside each pair of two counter-propagating plane modes of the trap. All such different pairs are decoupled from each other in the uniform BEC, and the covariance matrix becomes a diagonal matrix composed of the $(4 \times 4)$-blocks. So, the characteristic function factorizes into a product of the $(4 \times 4)$-determinants found in \cite{PRA2000} and the joint distribution manifests the squeezed two-mode fluctuations with correlations analogues to the ones in the Hong–Ou–Mandel effect of a two-photon interference in quantum optics.

Third, the $\sharp$P-hardness disappears in some exactly soluble or special cases when the Bogoliubov coupling matrix has a special or degenerate form such that the associated hafnians or permanents, defining the joint probability distribution in accord with the hafnian, Eq. (\ref{sm-HMT}), or permanent master theorems, are computable in polynomial time (e.g., via fully polynomial randomized approximation scheme \cite{Jerrum2004} or recursively, like permanents in \cite{Entropy2021}).
\\

\clearpage
\twocolumngrid


\begin{thebibliography}{99}
\bibitem{footnoteTrapSize} The size of the qubit well is of the order of the de Broglie wavelength, that is about $1\ {\mu}m$. The size of the entire multi-qubit BEC trap, say, in the 2D case is $\sim\sqrt{Q}\ {\mu}m$.
\bibitem{footnote1D} Another design could be based on a bunch of $Q$ quasi-one-dimensional BEC traps each supporting a preselected (say, via a properly designed Bragg reflection) longitudinal atomic mode and coupled to neighbours via a quantum tunnelling through a potential barrier of its wall. A cross-section profile of each 1D trap could support two, almost degenerate, transverse atomic modes with close lower energy levels constituting a qubit for each longitudinal mode. All higher-order transverse atomic modes could have much higher energy levels separated from the two qubit levels by a gap $\Delta E \gg T$.
\bibitem{footnoteDecoupledMiniband} At certain conditions, the $2^Q$ lower-miniband states are mainly decoupled from the continuum of excited states of the total infinite-size Hilbert space and constitute a finite-size Hilbert subspace of $Q$ interacting qubits. 
The quantity of interest is the joint probability distribution of fluctuations in their occupations in the thermal state $\hat{\rho}$ specified by the condensate wave function.
\bibitem{footnoteImaging} A single-shot optical imaging with a multi-detector recording could give information on the joint, simultaneous occupations of different cells/modes of the trap. Of course, particles in the condensate $\psi_0$, which is orthogonal to the excited states $\psi_k$, should not be countered. The cells/modes represent a complete set of the orthogonal excited states -- a basis. Importantly, the result for their joint occupation probability is universal relative to a choice of such a basis. A basis change amounts just to an additional unitary rotation, i.e., to a proper choice of the unitary $R$, Eq.~(\ref{BdG_transform}). Various coarse-grained measurements are also fully appropriate.
\end{thebibliography}

{}

\end{document}